\documentclass[
reprint,
superscriptaddress,
amsmath,
amssymb,
aps,
pra, 
longbibliography,
floatfix
]{revtex4-1}

\usepackage{graphicx}% Include figure files
\usepackage{xcolor}% colouring text

\usepackage{tabularx}% better tables
\usepackage{dcolumn} 
\newcolumntype{d}[1]{D{.}{.}{#1}}
\usepackage{booktabs}

\usepackage{comment}% comment multiple lines
\usepackage{enumitem}

\begin{document}

\title{Royal Society Inaugural Article\\
Perspective: Multiferroics Beyond Electric-Field Control of Magnetism}

\author{Nicola~A.\ Spaldin}
\affiliation{Department of Materials, ETH Zurich, CH-8093 Z\"{u}rich, Switzerland}

\date {\today}

\begin{abstract}
Multiferroic materials, with their combined and coupled magnetism and ferroelectricity, provide a playground for studying new physics and chemistry as well as a platform for development of novel devices and technologies. Based on my July 2017 Royal Society Inaugural Lecture, I review recent progress and propose future directions in the fundamentals and applications of multiferroics, with a focus on unanticipated developments outside of the core activity of electric-field control of magnetism.
\\
\\
\noindent
\color{red}{This is the Accepted Author Manuscript version of an article published in {\it  Proceedings of the Royal Society A} {\bf 476}, 20190542 (2020). \copyright 2020 Nicola A. Spaldin\\
\\
%\vspace*{0.5cm}
\noindent
The final authenticated version is available online at:
http://dx.doi.org/10.1098/rspa.2019.0542}

\end{abstract}

\maketitle

\section{Introduction}

The challenge of designing materials that combine multiple functionalities that tend not to coexist is intriguing from both fundamental and applications perspectives. As a scientist, understanding and then overcoming \textit{chemical contraindication} in which one type of crystal chemistry favors one property, but a second property requires a different type and arrangement of atoms, is an engaging research objective, often requiring identification of novel structure-property relationships. And as an engineer, a material that can perform more than one task is highly appealing for developing novel technologies. Examples of materials with sought-after contra-indicated functionalities include those with high permittivity combined with high permeability for antenna applications, insulating ferromagnets for high frequency devices, transparent conductors for displays, and efficient thermoelectric materials, with both high electrical and low thermal conductivity, to harness waste heat for power generation.

In particular, as information technologies consume an ever greater percentage of global energy resources, there is an urgent need for multifunctional materials that can reduce the energy demands of microelectronic devices. Particularly promising in this context is the class of materials known as magnetoelectric multiferroics, which combine the contra-indicated functionalities of ferromagnetism and ferroelectricity in the same phase. While there had been sporadic interest in magnetoelectric behavior starting in the 1950's \footnote{Indeed, the field of multiferroic materials emerged from a small but long-standing effort to develop {\it magnetoelectric} materials (see for example Ref.~\onlinecite{Freeman/Schmid:1975}), in which an applied electric field generates a linear magnetization and vice versa.}, progress was hampered by a scarcity of materials, until the combination of improved theoretical understanding \cite{Hill:2000} with modern synthesis and characterization techniques caused a boom in multiferroics research at the turn of this century.  The initial technological driver for this multiferroics renaissance was their potential for allowing electric-field control of magnetism \cite{Spaldin/Fiebig:2005,Spaldin/Ramesh:2008}, and the associated promise of entirely new, energy-efficient device architectures. There has been tremendous progress towards this goal and I direct the reader to recent reviews on this aspect \cite{Spaldin/Ramesh:2019}. In addition, as a ``side effect''  of the intense research effort on such magnetoelectric coupling in multiferroics, there have been many initially unexpected developments both in fundamentals of materials physics and in other application areas that result from their particular materials chemistry. These form the focus of this perspective. 

This article is organized as follows. We begin with a brief review of the chemical origin of the contra-indication between ferromagnetism and ferroelectricity, as well as routes to overcoming it. While this topic has been thoroughly discussed elsewhere, the focus here is on the mechanistic developments that are relevant for applications other than electric-field control of magnetism. Then follows a survey of applications, starting with a brief discussion of energy efficient devices that exploit electric-field control of magnetism, to contrast with bio-medical applications that rely on the reciprocal effect of magnetic-field control of electrical properties.  A discussion of the promising photovoltaic and photocatalytic properties that result from the unusual combination of {\it chemistries} required for multiferroicity is followed by applications in cosmology and high-energy physics that result from their unusual combination of {\it physical} properties. Finally, we come full circle and show how the exotic physics suggests new energy-efficient device architectures outside of the electric-field control of magnetism paradigm. The article closes with some thoughts on future promising research and technology directions.

\section{The contra-indication between ferroelectricity and ferromagnetism}

A ferroelectric is defined to be a material with a spontaneous electric polarization (that is a dipole moment per unit volume) that is switchable by an applied electric field. To achieve such behavior, the anions and cations in a material have to be off-centered relative to each other (to form the electric dipole) but not so rigidly that this off-centering can not be reversed by the field. The definition excludes all metals, because the electrons in a metal screen electric fields and prevent them from exerting forces on the ions, and favors chemistries with polarizable, field-responsive bonds. Compounds of first-row transition metals with oxygen are particularly suitable, since they fulfill these criteria, as well as being earth-abundant and environmentally benign. An additional factor determines whether the lowest energy arrangement of the ions in a transition-metal oxide is polar and ferroelectric, or centrosymmetric and non-ferroelectric: Since the electrons on isolated ions tend to be highly symmetrical or even spherical, their short-range Coulomb repulsion with electrons on neighboring ions is minimized if the ions are packed in a symmetric, non-ferroelectric fashion. On the other-hand, energy-lowering chemical bond formation is strongest when pairs of anions and cations are close to each other forming local electric dipoles and a ferroelectric state \cite{Burdett:1981,Rondinelli/Eidelson/Spaldin:2009}. The balance between these competing behaviors, collectively known as the second-order Jahn-Teller effect, tends to be tipped towards ferroelectricity for the combination of oxygen O$^{2-}$ anions with transition metal cations that have {\it empty} valence $d$-electron manifolds, but towards the centrosymmetric state when the valence $d$ orbitals are partially filled. Herein lies the chemical incompatibility between ferroelectricity and magnetism: An empty $d^0$ electronic structures favors ferroelectricity, but can not host magnetism since it has no unpaired electrons to provide a magnetic moment.

\section{Ways around the contraindication}
\label{Ways_around}

Many mechanisms for overcoming the contra-indication between the $d^0$-ness that favors ferroelectricity and the $d^n$ occupation needed for magnetism have been identified over the last decades. Here we review those that are relevant for the applications discussed below, noting that a single-phase material with large and robust magnetization and polarization, as well as strong coupling between them at room temperature remains to be identified. For a more complete review, including of important examples such as magnetically induced ferroelectricity \cite{Newnham_et_al:1978,Kimura_et_al_Nature:2003} that are not covered here, see for example Refs.~\onlinecite{Spaldin/Ramesh/Cheong:2010,Spaldin/Ramesh:2019}. 

\begin{figure}
\centering
\includegraphics[scale=0.6]{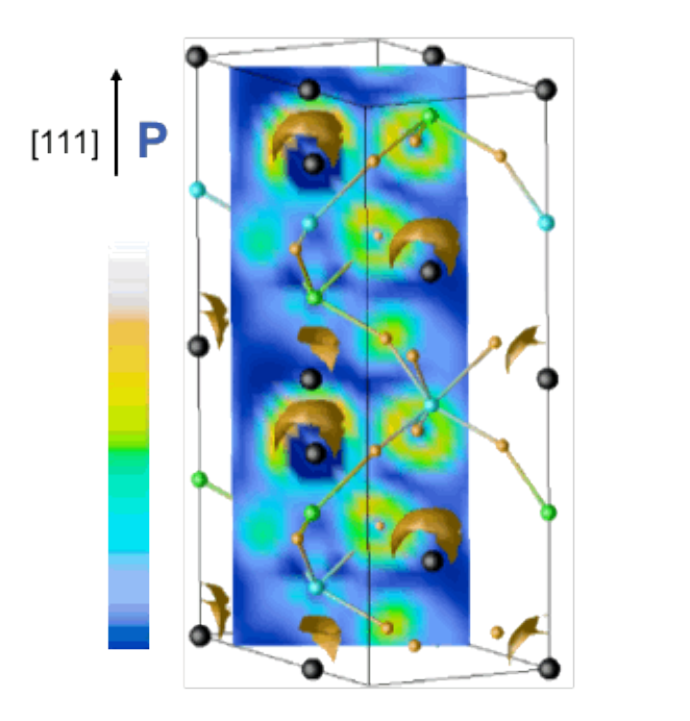}
\caption{Calculated electron localization function for multiferroic BiFeO$_3$ indicating the ferroelectric polarization, P, caused by the stereochemically active lone pairs of electrons (yellow umbrella-shaped lobes) on the Bi ions (black spheres). The color scale runs from highly localized (white) to completely delocalized (dark blue). Fe and O atoms are shown in green and yellow, respectively. Reproduced with permission from Ref.~\onlinecite{Spaldin_MRS:2017}.}
\label{fig:bifeo3_lonepair}
\end{figure}

Conceptually the most straightforward way to circumvent the contra-indication is to combine two separate materials -- one of them ferroelectric and one ferro (or ferri)magnetic -- in a composite. This approach has met with success over multiple length scales, from nanocomposites \cite{Zheng_et_al:2004} and ultra-thin film heterostructures \cite{Mundy_et_al:2016} to macroscopic laminates \cite{Vaz_et_al:2010}, and provides the materials of choice for the biomedical applications described below. A more sophisticated approach exploits the fact that materials can contain cations of more than one type -- the A- and B-site cations in the perovskite structure for example -- and then uses one of them to drive the ferroelectricity and the other the magnetism. Perhaps the leading example of this approach is the most widely studied multiferroic, bismuth ferrite, BiFeO$_3$ \cite{Wang_et_al:2003}, in which the $d^5$ Fe$^{3+}$ ions provide the magnetism and the ferroelectric distortion is driven by the stereochemical activity of the Bi$^{3+}$ lone pair of electrons (Fig.~\ref{fig:bifeo3_lonepair}) \cite{Hill/Rabe:1999, Seshadri/Hill:2001}. In addition to its leading role in electric-field control of magnetism, the presence of partially filled iron $3d$ orbitals cause the states around the band gap to be a mix of O $2p$ and Fe $3d$, which is relevant for applications in catalysis and water purification. The converse approach -- a $d^0$ B-site transition metal and a magnetic A site -- is exemplified by modifications of EuTiO$_3$. While not ferroelectric under ambient conditions due to the small size of the Eu$^{2+}$ ion providing insufficient space for the Ti$^{4+}$ to off center, both tensile strain \cite{Lee_et_al:2010} and negative chemical pressure \cite{Rushchanskii_et_al:2010} have been shown to induce ferroelectricity. The $f^7$ Eu$^{2+}$ ions have large spin magnetic moments, which do not interact strongly because of the highly localized nature of the $f$ electrons. While the resulting low  magnetic ordering temperature is unfavorable for device applications, its combination with the substantial and robust ferroelectricity from the $d^0$ Ti$^{4+}$ ion has enabled the search for subtle fundamental physics that relies on time-reversal and space-inversion symmetry breaking at low temperature \cite{Rushchanskii_et_al:2010}. 

\begin{figure}
\centering
\includegraphics[scale=0.45]{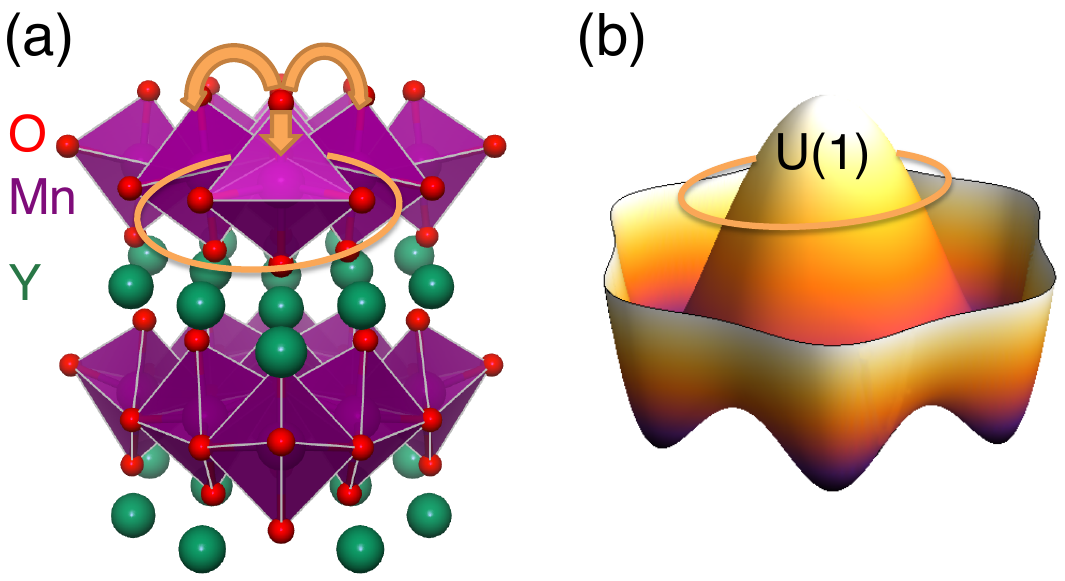}
\caption{(a). Crystal structure of multiferroic yttrium manganite, YMnO$_3$. The structure consists of layers of Y$^{3+}$ ions (green) separating layers of MnO$_5$ trigonal bipyramids (purple), in which the Mn$^{3+}$ ions form a triangular lattice linked by the oxygen ions (red) at the corners of the bipyramids. The ferroelectricity emerges from the tilting of the MnO$_5$ bipyramids (orange arrows) which lowers the energy by the same amount for any tilt angle when the tilts are small (orange circle). The Y$^{3+}$ ions shift in response to the tilting, causing the ferroelectric polarization. (b) Detailed symmetry analysis shows that the potential describing the phase transition has the form of the ``Mexican hat'' shown in the Figure, in which the distance from the peak represents the magnitude of the tilting and the angle around the hat its orientation. For small tilts, the angle degeneracy is reflected in the U(1) symmetry; for larger tilts, six angles are favored, giving rise to six ferroelectric domains. Reproduced with permission from Ref.~\onlinecite{Griffin/Spaldin:2017}.}
\label{fig:ymno3}
\end{figure}

The search for ferroelectric mechanisms that are compatible with the existence of magnetism has unearthed a number of new driving forces for off-center distortion other than the usual covalent bond formation that occurs with the $d^0$ or lone-pair cations mentioned above. Of these, so-called geometric ferroelectricity \cite{vanAken_et_al:2004} is relevant for two of the applications that we will discuss below. In geometric ferroelectrics, the rotational distortions of the transition-metal / oxygen polyhedra, which are common in many ceramics but are usually non-polar, introduce a polarization because of unusual lattice connectivity \cite{Ederer/Spaldin_3:2006}. In the hexagonal manganites, of which YMnO$_3$ is the prototype (Fig.~\ref{fig:ymno3} (a)), the ferroelectricity is both geometric and improper \cite{Fennie/Rabe_YMO:2005,Artyukhin_et_al:2014}, meaning that the polarization arises from its coupling to the primary, zero-polarization rotational distortion, leading to highly unconventional domain formation \cite{Griffin_et_al:2012} and resulting properties \cite{Meier_et_al:2012}. 

\section{Applications of multiferroics}

\subsection{Energy-efficient devices I: Electric-field control of magnetism}

As stated above, the original technology driver for the study of multiferroics was the tantalizing possibility of controlling magnetism using electric, rather than magnetic, fields. Here, we give a brief overview of the progress in this direction with the purpose of illustrating the evolution and time line from fundamental science in the laboratory to prototype devices; for a more detailed review see Ref.~\onlinecite{Spaldin/Ramesh:2019}. The turn of the 21st Century marks the beginning of the ``modern era'' of multiferroics, when the basic principles of the contra-indication between ferroelectricity and magnetism were articulated and candidate mechanisms, particularly the incorporation of stereochmically active lone pairs into magnetic materials, were proposed \cite{Hill:2000}. Three years later, the first demonstration of large and robust room-temperature ferroelectricity was made in thin films of multiferroic BiFeO$_3$ \cite{Wang_et_al:2003}. The first breakthrough in the electric-field control of magnetism using multiferroics followed in 2006 in the same material, with the demonstration of the reorientation of the antiferromagnetism using an electric field \cite{Ederer/Spaldin:2005,Zhao_et_al:2006}. Since the lattice deformation associated with the ferroelectric polarization determines the plane in which the Fe magnetic moments are oriented,  electric-field reorientation of the polarization in turn reorients this magnetic easy plane. This finding generated considerable excitement, and was listed by Science magazine as one of their up-and-coming {\it Areas to Watch} \cite{Science:2007}. Of course, more appealing is the electric-field control of a net magnetization. An important step in this direction was the 2010 demonstration that electric-field reorientation of the antiferromagnetic domain orientation in magnetoelectric Cr$_2$O$_3$ affects the exchange-bias coupling to a ferromagnetic overlayer \cite{He_et_al:2010}. Switching of the ferromagnetic layer was subsequently achieved four years later by coupling a layer of ferromagnetic metallic CoFe to BiFeO$_3$ in a thin-film heterostructure \cite{Heron_et_al:2014}. This progression culminated in 2018 with the demonstration of a prototype {\it magnetoelectric spin-orbit logic} (MESO) device that promises substantially lower energy consumption than existing architectures \cite{Manipatruni_et_al:2018}. A challenge for multiferroics researchers is to reduce the voltage required to switch the ferroelectric state of the magnetoelectric component to around 100 mV from the current value of $\sim$ 5 V. At the same time, for the MESO device to be competitive, the voltage output from the spin-orbit component should be increased to hundreds of mV from its current  $\mu$V values \cite{Spaldin/Ramesh:2019}. This combination of improvements is predicted to yield memory bits or logic elements that could be operated with 1 aJ of energy per bit, which would be technologically transformative.

\subsection{Bio-medical applications: Magnetic-field control of electrical properties}

While magnetic-field control of electrical properties had not been a primary driver of multiferroics research, the fact that magnetic fields can be applied remotely, without the need for electrodes, is appealing for {\it in vivo} applications. Astonishing progress has been made in recent years in exploring multiferroics and magnetoelectrics for two bio-medical application areas: magnetically assisted in vivo targeted drug delivery, and enhanced scaffolds for tissue engineering. 

For the drug delivery application, the approach is as follows: A magnetic field is first used to guide small multiferroic particles carrying drugs to the chosen site in the body, via the direct magnetic interaction with the ferromagnetic component. Then the coupled magnetic / ferroelectric / ferroelastic property of multiferroics allows application of an oscillating magnetic field to change both the electric polarization (and hence surface charge) and shape of the particles. This in turn breaks the drug-nanoparticle bond, triggering the release of the drug molecules directly at the chosen site. 

Material demonstrations of such particulate micromachines include core-shell nanoparticles composed of  ferromagnetic and magnetostrictive FeGa cores surrounded by ferroelectric and piezoelectric polymeric shells \cite{Chen_et_al:2017}, piezoelectric polymeric matrices containing magnetic nanoparticles \cite{Mushtaq_et_al:2019,Chen_et_al_MatHoriz:2019}, and ceramic ferrimagnetic CoFe$_2$O$_4$ / ferroelectric BaTiO$_3$ composite nanoparticles \cite{Nair_et_al:2013,Chen_et_al:2016}.
Mouse tests have shown that the latter can be magnetically guided across the blood-brain barrier, and that they do not cause toxicity or behavioral impairment \cite{Kaushik_et_al:2016}, suggesting promise in treating traumatic neuronal injuries or degenerative diseases via delivery of drugs or neural progenitor cells and through cell stimulation \cite{Chen_et_al:2017}. 

Restoration of tissue such as muscle or bone after disease or injury is an increasingly relevant challenge in our aging society, and tissue engineering is an active area of fundamental and translational research. As a proxy for the extra-cellular matrix of the native tissue, artificial scaffolds are often used to facilitate the migration and attachment of cells as well as the diffusion of cell nutrients and waste products. Since cells are known to proliferate more rapidly when subject to mechanical or electrical stimulation, piezoelectrically active materials are promising choices as novel scaffolds for tissue engineering. They are particularly relevant as scaffolds for bone regrowth, since bone itself is piezoelectric, and the electromechanical stimulation produced by day-to-day activities is believed to be relevant for natural bone regeneration. Again, multiferroics offer the additional advantage that mechanical and electrical stimulation can be achieved remotely from outside of the body using magnetic fields. In particular, magnetoelectric composites of piezomagnetic Terfenol-D with piezoelectric poly(vinylidene fluoride-co-trifluoroethylene) have been shown to provide mechanical and electrical stimuli to bone-synthesizing cells (so-called osteoblasts) on application of a magnetic field \cite{Ribeiro_et_al:2016}. In another study, the additional bone-like feature of porosity was introduced by preparing inverse-opal scaffolds from biodegradable poly(l-lactic acid) combined with cobalt ferrite/bismuth ferrite magnetoelectric nanoparticles. A substantial increase in cell proliferation was found on application of a magnetic field \cite{Mushtaq_et_al:2019/3}.

A challenge for all biomedical applications is that the materials must be biocompatible and non-cytotoxic. In addition, long-term stability is required for structures that are designed to be implanted for a long period in the body. Perhaps a greater challenge is faced in designing systems that are designed to be eliminated, since ferroelectric and / or ferromagnetic properties are often not maintained at the small particle sizes (typically below 10 nm) that can be excreted renally.

\subsection{Clean energy and the environment: Photovoltaics and photocatalysis} 

The technological focus at the start of the multiferroics renaissance was on their promise in designing microelectronic devices that use less energy. In an unexpected complement to this original direction, multiferroics are now showing promise  as materials for \textit{harvesting} energy, specifically from sunlight. In photovoltaic and photocatalytic processes, a semiconductor absorbs photons from sunlight and converts them into electron-hole pairs which then generate electrical current or initiate a chemical reaction. Ideally, the semiconductor will have a low enough band gap to capture the full solar spectrum, and some mechanism for preventing the recombination of the electrons and holes. Multiferroics are ideally suited for both tasks: The internal electric fields associated with the presence of ferroelectricity cause electrons and holes to move in opposite directions, and the presence of transition-metal $d$ states in the region of the Fermi energy tends to reduce the band gap enabling capture of low-energy solar phonons. In addition, there are possibilities of both electric-field control and magnetic-field control of properties such as the photocurrent direction. 

While in theory, the photovoltaic response in multiferroics is more complex than in conventional ferroelectrics because the light acts on the coupled magnetic and ferroelectric properties \cite{Mettout/Toledano:2019}, in practice the process appears to be limited by extrinsic factors, and optimization of photovoltaic efficiency is a rich area for empirical research. Following the first report of a photodiode effect in crystals of BiFeO$_3$ \cite{Choi_et_al:2009}, most studies have focused on this material, since it is easy to synthesize, has a direct band gap of around 2 eV, and one of the largest known ferroelectric polarizations ($\sim$90 $\mu$C/cm$^2$). All-oxide heterostructures with SrRuO$_3$ bottom electrodes, BiFeO$_3$ active layers, and transparent tin-doped indium oxide top electrodes \cite{Yang_et_al:BFO:2009} are particularly promising, offering open-circuit voltages of $\sim0.8$ V and external quantum efficiencies up to 10 \%. The power conversion efficiency is improved by A-site or B-site doping and by increase of oxygen vacancy concentration \cite{Mocherla_et_al:2013}, likely as a result of reducing the band gap, and is sensitive to microstructural characteristics, such as domain structure, strain and film thickness or nanoparticle size \cite{Mocherla_et_al:2013}. For a recent review and a summary of the performance of various chemistries and morphologies see Ref.~\onlinecite{Chen_et_al:2019}.

In addition to the requirements listed above for photovoltaics, a photocatalytic material must be able to adsorb the starting molecules on the surface, provide a pathway for a chemical reaction, and then release the product molecules into the surroundings. The combination of the polarizability of the oxide ion with the localized transition-metal $3d$ states in the region of the Fermi level means that transition-metal oxides often have favorable surface properties for  adsorption and reactivity of molecules. In multiferroics, the internal electric fields associated with the presence of ferroelectricity may enhance these properties further, as well as offering the possibility of using electric fields to ``shake'' the surface through the piezoelectric effect to promote release of the adsorbed species. This possibility has motivated both fundamental studies of the functionalities of multiferroic surfaces, largely in nanostructured form, as well as feasibility tests of various catalytic applications. As in the case of photovoltaics, the material of choice to date has been bismuth ferrite, owing to its ease of production, cheap and widely available constituents, stability and absence of toxicity, all of which are important for the target applications, as well as its low band gap, large polarization and substantial piezoelectric response. 

An environmentally relevant application of photocatalysis is water purification through the catalytic degradation of high molecular weight organic molecules such as contaminant dyes into smaller molecules such as CO$_2$ or H$_2$O.  BiFeO$_3$ nanosheets, nanowires  and nanoparticles have been demonstrated to photo-degrade large organic dyes such as methyl orange \cite{Xian_et_al:2011} and rhodamine B \cite{Soltani/Entezari:2013} in the presence of sunlight, with mechanical vibrations increasing the efficiency likely due to the internal piezoelectric fields' promoting charge separation \cite{Mushtaq_et_al:2018}. Intriguingly, magnetoelectric cobalt ferrite / bismuth ferrite core–shell nanoparticles were shown to catalyze the breakdown of synthetic dyes and pharmaceuticals in the presence of an AC magnetic field, even without sunlight \cite{Mushtaq_et_al:2019/2}, highlighting a need for improved understanding of the interplay between applied fields, the surface electronic structure, the defect chemistry and microstructure, and the environment. A helfpul review of synthesis methods for BiFeO$_3$ nanoparticles, and the factors that affect their efficiency in environmental mediation, can be found in Ref.~\onlinecite{Gao_et_al:2015}. Development of processes to control the ferroelectric domains, which in turn influence surface reactivity, in multiferroic nanoparticles remains a challenge.

Finally, I mention some other potential environmentally relevant applications that have been explored for bismuth ferrite, that depend on its multiferroic behavior but are not related to the electric-field control of magnetism. Composites of organic pectin with BiFeO$_3$ nanoflakes have been explored for electrodes in rechargeable Na-ion batteries and shown to offer favorable rate and cycling performance \cite{Sun_et_al:2018}. A high Seebeck coefficient has been reported, suggesting possible thermoelectric relevance \cite{Yokota/Aoyagi/Gomi:2013}, as well as photostriction \cite{Kundys_et_al:2010}, photodetection \cite{Qi_et_al:2018}, gas sensing behavior \cite{Waghmare_et_al:2012,Dong_et_al:2015}, and resistive switching, with possible application in neuromorphic computing \cite{Kolhatkar_et_al:2018}. Finally, (Bi,Sr)FeO$_3$, with its high concentration of oxygen vacancies combined with high vacancy mobility, likely due to its large polarizability, shows promise for solid-oxide fuel cells \cite{Wedig_et_al:2011}.  Further studies are needed in all these directions to understand the mechanisms and assess feasibility.

\subsection{Cosmology and high-energy physics}

The most personally enriching aspect of my own research on multiferroics has been the surprising finding that their properties can have implications in other branches of often rather fundamental physics~\cite{Spaldin_NRM:2017}. In many cases this is a result of their unusual combination of symmetry breakings -- their ferroelectricity means that they break space-inversion symmetry, and their magnetic ordering results in time-reversal symmetry breaking -- and so they provide an analogue to more fundamental processes in space/time with the same symmetry properties. Here I describe my three favorite examples, and discuss some possible future directions.

\subsubsection{Search for the electron electric dipole moment.} 
For materials scientists, both the charge and the spin of the electron are familiar properties that are essential for the behavior of materials, the former being responsible for example for ionic bonding and the latter for magnetism. We tend to think less often about the {\it electric dipole moment of the electron}, which, while proposed to be very small, must be non-zero according to all descriptions of the universe beyond the simplest Standard Model. For a multiferroicist, the electron electric dipole moment is particularly intriguing, since by symmetry it must have the same axis of orientation as the electron's magnetic moment, meaning that when the magnetic dipole moment of an electron is reversed by a magnetic field, its electric dipole moment must reverse too (and vice versa), showing ideal multiferroic behavior! This correspondence makes multiferroics a suitable platform for searching for the electron electric dipole moment: In an applied electric field $\vec{E}$, the usual Stark effect will cause an electron with its electric dipole moment $\vec{d}$ parallel to the field to have a lower energy by an amount $2\vec{d.E}$ than one oriented antiparallel to the field. As a result, at any finite temperature, more electrons will align with their electric dipoles parallel to the field than opposite to it, and a corresponding imbalance in the total magnetization must occur that should be detectable by sensitive magnetometry measurements \cite{Lamoreaux:2002}. Since the effect is tiny, however, a material with a large ferroelectric moment to amplify the effect of the external electric field is desirable, combined with unpaired electrons, with corresponding magnetic moments, so that their electric dipoles don't cancel out \cite{Sushkov/Eckel/Lamoreaux:2010}.  Multiferroic (Eu,Ba)TiO$_3$, with its large ferroelectric polarization and suppressed magnetic order \cite{Rushchanskii_et_al:2010},  has enabled the highest-precision solid-state search to date for the electron dipole moment, giving an upper bound on its possible value of $6.05 \times 10^{-25}$ ecm \cite{Eckel/Sushkov/Lamoreaux:2012}. Higher accuracy was thwarted by hysteretic heating during the switching process, which in turn limited control of the cryogenics, and circumventing this issue will remain a challenge in future solid-state searches.

The existence and magnitude of the electron electric dipole moment have profound implications for fundamental physics (for a review see Ref.~\onlinecite{Fortson/Sandars/Barr:2003}.) Since reversing time reverses the electron's magnetic moment but not its electric dipole, the presence of an electron electric dipole moment indicates that time-reversal symmetry is broken in the universe. The product of charge conjugation (C), parity inversion (P) and time reversal (T) is known to be an invariant, however, and so the breaking of time-reversal symmetry implies a breaking of CP symmetry.  Since various fundamental theories incorporate this property in different ways, they in turn make different, well-defined predictions about the size of the electron electric dipole moment. 

The most precise value to date, obtained from optical measurements on molecular ThO by the ACME collaboration, is a remarkable upper bound of $1.1 \times 10^{-29}$ ecm \cite{Baron_et_al:2014,Andreev_et_al:2018}. The deviation from spherical that such a small dipole causes in the electron's charge is the same as that caused by removing a couple of nanometers of material from the top of a sphere the size of the earth and adding it to the bottom! As a result of this ground-breaking experiment, a number of leading beyond-standard-model theories had to be eliminated, including split supersymmetry and spin-10 grand unified theory. The value also suggests that any beyond-standard-model particles must be more massive than previously proposed, reducing the chances that they can be detected by accelerators such as the Large Hadron Collider.

\subsubsection{Magnetoelectric monopoles and beyond.}

\begin{figure}
\centering
\includegraphics[scale=0.4]{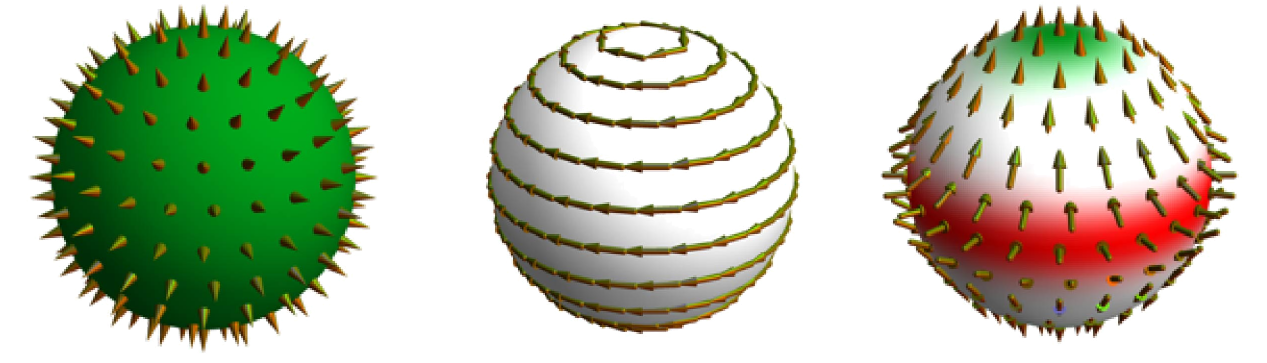}
\caption{Representation of (left to right) a magnetoelectric monopole, the $z$ component of the toroidal moment and the $z^2$ component of the quadrupole moment. The arrows represent the orientation of the magnetic moment at each point. Reproduced with permission from Ref.~\onlinecite{Spaldin_et_al:2013}. Copyright (2013) by the American Physical Society.}
\label{fig:Hedgehog}
\end{figure}

As discussed in the Introduction, many multiferroics exhibit a linear magnetoelectric effect, in which a magnetic field induces a polarization linearly proportional to its strength, and reciprocally an electric field induces a linear magnetization. It has been pointed out that the linear magnetoelectric response can be recast in terms of so-called \textit{magnetoelectric multipoles}, which form the second-order term in the multipole expansion of the energy of a general magnetization in a general inhomogeneous magnetic field~\cite{Spaldin/Fiebig/Mostovoy:2008}. It is helpful to decompose the magnetoelectric multipoles, into monopolar, toroidal and quadrupolar components, illustrated in Figure~\ref{fig:Hedgehog},  which couple to the divergence, curl and gradients of the magnetic field respectively. Materials with a net magnetoelectric monopole have a diagonal isotropic component in their linear magnetoelectric response; toroidal moments and quadrupoles yield antisymmetric and traceless responses respectively.

The monopolar contribution to the magnetoelectric response has an intriguing connection to high-energy physics via a proposed elementary particle, known as the {\it axion}, a pseudoscalar that was introduced to ameliorate the strong-CP problem of quantum chromodynamics \cite{Wilczek:1978,Weinberg:1978}. Axions are generated by adding a term of the form ${\cal L} \propto a \mathbf{E.B}$ to the usual Maxwell Lagrangian \cite{Wilczek:1987}, and so break both time-reversal and space-inversion symmetries in the same way as the magnetoelectric monopole. In particular, they exhibit a diagonal linear magnetoelectric response, acquiring an electric dipole moment parallel to a uniform applied magnetic field and vice versa.

Other exotic behaviors proposed for these magnetoelectric monopoles, include a divergent magnetic field generated by an electric charge above a slab of magnetoelectric material \cite{Meier_et_al:2019}, and a spontaneous magnetoelectric Hall effect in the absence of an applied magnetic field \cite{Khomskii:2013}. Hints of these behaviors have been seen in muon spin rotation measurements \cite{Meier_et_al:2019,Dehn_et_al:2019} but they still await unambiguous confirmation. Possible ``monopolotronic'' applications, for example in data storage or processing, are as yet unexplored.

\subsubsection{Cosmic string formation in the early universe.} 

The ferroelectric and ferromagnetic states in multiferroics, in which the electric or magnetic dipoles are aligned in an ordered fashion, are often obtained by cooling through phase transitions from higher symmetry dipole-disordered paraelectric or paramagnetic phases, respectively. Such spontaneous symmetry-breaking phase transitions are not unique to multiferroics, but occur in a wide range of physical systems, from low-energy cold atom experiments, through superconducting transitions in condensed matter, to Higgs boson formation at the Large Hadron Collider \cite{Baumann_et_al:2010, Sadler_et_al:2006,Englert/Brout:1964,Higgs:1964,Guralnik_et_al:1964}. An intriguing incidence of spontaneous symmetry breaking is proposed by the `standard model' of cosmology, which suggests that expansion and cooling of the early universe drove a symmetry-lowering phase transition -- called the Grand Unification Transition -- in the primordial vacuum at $\sim 10^{-37}$ seconds after the Big Bang. The symmetry breaking in this case is described by a so-called `Mexican-hat' potential, in which the order parameter, which is zero in the disordered phase at $T > T_{C}$, acquires a non-zero value that is independent of its orientation below $T=T_{C}$. In such a phase transition, one-dimensional topologically protected defects consisting of the higher symmetry vacuum phase should be trapped within the lower-symmetry phase, and their concentration should depend on the rate at which the phase transition proceeds according to well-defined scaling laws known as Kibble-Zurek scaling \cite{Kibble:1976,Zurek:1985}. The search for signatures of these defects, known as cosmic strings, is an exciting challenge in observational cosmology.

The Mexican-hat potential has a continuous symmetry, in that it is smooth as the angle around the peak is varied. Therefore it does not usually describe structural phase transitions in solids, which, because they consist of separate, individual atoms, have discrete symmetries (displacing the atoms in different directions yields different energy changes). It has been recently shown, however, that the unusual improper ferroelectric phase transition in the multiferroic hexagonal manganites described in Section~\ref{Ways_around} has a quasi-continuous symmetry, many aspects of which are well described by a Mexican-hat-like potential (Fig.~\ref{fig:ymno3} (b)) \cite{Artyukhin_et_al:2014, Griffin_et_al:2012}. As a result, simple laboratory experiments on the details of the phase transition in the hexagonal manganites can be used to shed light on the behavior of their more esoteric counterparts. In particular, the hexagonal manganites have been used as a model system to simulate early-universe cosmic string formation in the laboratory. 

\begin{figure}
\centering
\includegraphics[scale=1.2]{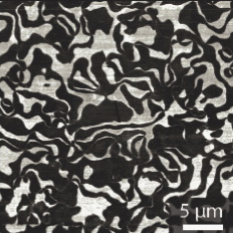}
\caption{Distribution of ferroelectric domains in YMnO$_3$ samples after quenching through the ferroelectric phase transition at a rate of 195 K/min. The image was obtained using piezoforce microscopy on 30 x 30 $\mu$m$^2$ area. Black and white regions indicate opposite orientations of the ferroelectric polarization which is pointing out of the plane of the image. The six-fold intersection points at the surface correspond to line defects in three dimensions. Reproduced from Ref.~\onlinecite{Griffin_et_al:2012}.}
\label{fig:YMnO3_domains}
\end{figure}

Figure~\ref{fig:YMnO3_domains} shows a cross-section of the ferroelectric domain structure of YMnO$_3$ measured using a technique called piezoforce microscopy, with the opposite orientations of ferroelectric polarization indicated by the the black and white regions \cite{Griffin_et_al:2012}. In this surface cross section, the alternating domains appear to intersect at points, which correspond to one-dimensional lines in three dimensions. Intriguingly, and unlike in conventional ferroelectrics, the lines are insensitive to an applied electric field, and the domain structure can not be converted to a single domain. The hexagonal manganites therefore confirm the first  mathematical requirement for the behavior of a phase transition described by a Mexican-hat potential, that is the formation of one-dimensional topologically protected defects. A second requirement is that these one-dimensional topologically protected defects are composed of lines of the high temperature structure, trapped within the low-symmetry phase \cite{Kibble:2003}; this has been confirmed using high-resolution transmission electron microscopy \cite{Zhang_et_al:2013}. Finally, the detailed predictions of Kibble-Zurek scaling can be tested directly, by cooling crystals of the hexagonal manganites at different rates through the phase transition, and counting the number of domain intersections that form. Remarkably, the measured density of domain intersections is well described by Kibble-Zurek scaling over a very wide range of cooling rates \cite{Griffin_et_al:2012}. An unexpected dependence on the identity of the A-site cation as well as a turn around at high cooling rates hint at additional as yet unexplained physics \cite{Meier_et_al:2017}. 

In other directions, detailed measurements of the local structure above and below the phase transition in the hexagonal manganites point to an unconventional continuous structural order-disorder behavior associated with the Mexican-hat potential physics \cite{Skjaervo_et_al:2019}. And on a more mundane, practical level, the electronic structure associated with their unusual five-fold manganese coordination leads to a symmetry-allowed optical transition in the red region of the spectrum. As a result they can be engineered to have an intense blue color, making them ideal as blue pigments in paint \cite{Smith_et_al:2009}. Whether the hexagonal manganites have anything to say about Higgs boson formation remains an open question that is being actively pursued \cite{Meier_et_al_Higgs:2019}.

\subsection{Energy efficient devices II: Domain wall functionality}

The unusual nature of the domain formation in the multiferroic hexagonal manganites that we discussed in the previous subsection results in properties that bring us full circle back to the discussion at the opening of this article: The need for materials that enable novel device architectures, which are more energy efficient than our existing technologies. 

\begin{figure}
\centering
\includegraphics[scale=0.6]{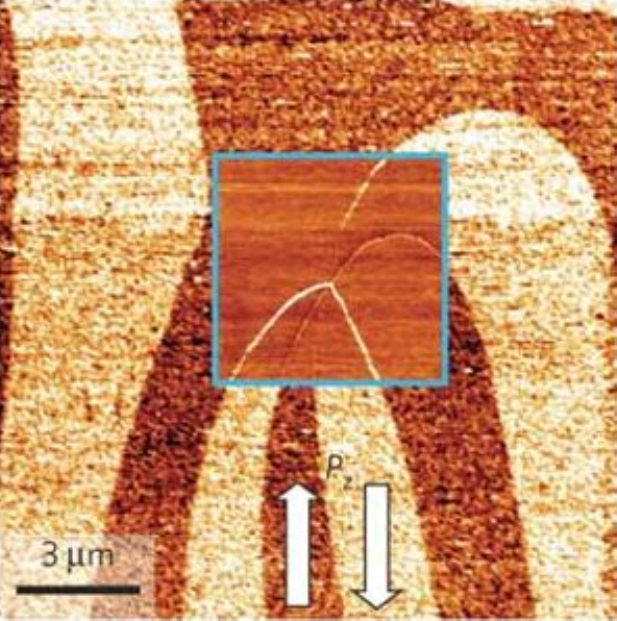}
\caption{The large image shows ferroelectric domains, measured using piezoforce microscopy, of hexagonal ErMnO$_3$ with its polarization orientation in the plane of the sample, with arrows indicating the direction of polarization. The inset shows a conducting atomic-force microscopy image acquired at the same position. Domain walls appear as lines of different brightness on an otherwise homogeneous background, with the tail-to-tail walls brightest, reflecting their enhanced conductance. Reproduced from Ref.~\onlinecite{Meier_et_al:2012}.}
\label{fig:YMnO3_DW}
\end{figure}

Figure~\ref{fig:YMnO3_DW} shows a similar domain structure to that shown in Fig.~\ref{fig:YMnO3_domains} for YMnO$_3$, this time for the related material ErMnO$_3$ and with the polarization (indicated by the white arrows) in the plane of the sample surface. The image shows data measured using both piezoforce microscopy and a technique called conducting atomic force microscopy, which provides a measure of local conductivity at the surface of the sample \cite{Meier_et_al:2012}.  The first unusual point to note is the existence of domain walls connecting regions of polarization that point directly towards or away from each other, called head-to-head and tail-to-tail walls respectively. Such domain walls are rare in conventional ferroelectrics since they are highly electrostatically unfavorable, requiring some modification of the system to avoid a divergence in the electrostatic potential. In the hexagonal manganites they occur because the primary order parameter, which sets the domain structure, is the zero polarization tilting of the MnO$_5$ polyhedra, and once the domains are formed the topological protection prevents them from reaching a lower energy solution. In the case of the hexagonal manganites, the unfavorable electrostatics is ameliorated by a transfer of charge that results in a change in the conductivity. We see in Figure~\ref{fig:YMnO3_DW} that at regions of tail-to-tail ferroelectric domain walls there is enhanced conductivity, whereas at head-to-head walls the conductivity is suppressed. In contrast, walls lying parallel and antiparallel to the polarization orientation have the same conductivity as the interior of the domains. The changes in conductivity are consistent with the native $p$-type nature of the hexagonal manganites: Free holes migrate to the tail-to-tail walls where their positive charge both screens the divergence of the electrostatic potential and provides the conductivity. There are no negatively charged free carriers to screen the polarization discontinuities at the head-to-head walls, which are compensated instead by depletion of holes, leaving them with a suppressed conductivity. There is no build up or depletion of charge at the parallel walls, which are therefore referred to as neutral domain walls.

The ferroelectric domain walls in the hexagonal manganites are just a few angstroms thick, and can be moved around using small electric fields. The fact that they carry an associated, orientation-dependent conductivity suggests entirely new device concepts that could enable novel applications in storage or sensing. Challenges to realizing such devices include controlling the motion and position of the domain walls, as well as fast detection of the conduction state.

\section{Summary and Outlook}

\begin{figure}
\centering
\vspace*{11pt}
\includegraphics[scale=0.1]{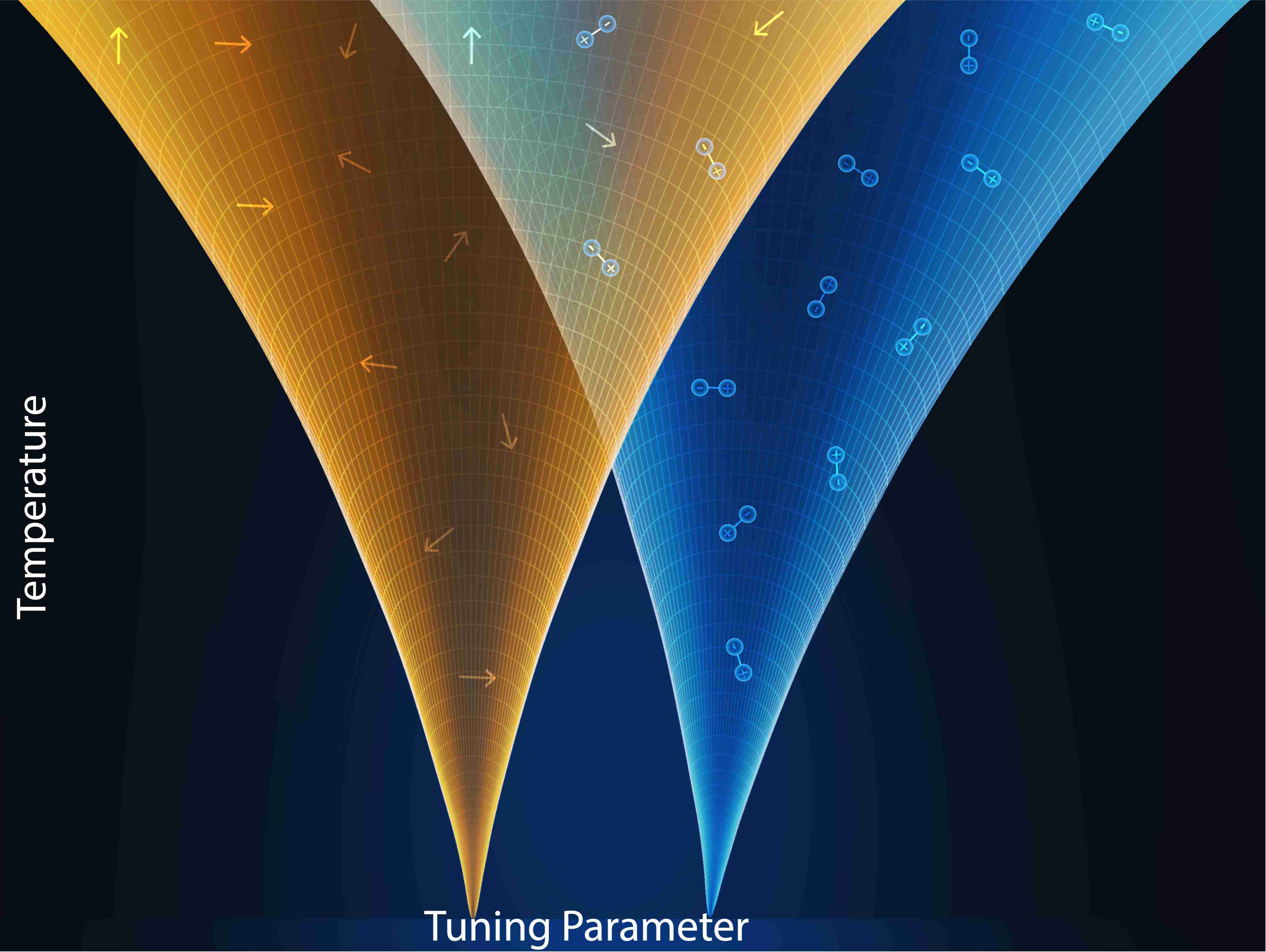}
\caption{Cartoon of the exotic behavior that persists to finite temperature (vertical axis) in the quantum critical fans above zero-temperature quantum phase transitions (at the points of the fans). The horizontal axis represents a variable other than temperature, for example pressure or applied field. When both magnetic (arrows) and ferroelectric (dumbbells) quantum critical points coexist and interact, additional novel physics is anticipated. Figure courtesy of Bara Krautz.}
\label{fig:MFQC}
\end{figure}

In summary, we have seen how the relevance of multiferroic materials now vastly exceeds the applications envisaged at the start of the multiferroics renaissance at the turn of this century. While there has indeed been tremendous progress on the original goal of electric-field control of magnetism and associated microelectronic devices, devices based on previously unimaginable phenomena such as conducting domain walls with polarization discontinuities suggest even greater transformative potential for new technology architectures. Applications that exploit the unique combination of chemical attributes required for ferroelectricity and magnetism to coexist, ranging from photocatalysis to drug delivery to blue pigments, offer tremendous potential. And the basic physics of combined time-reversal and space-inversion symmetry breaking mean that multiferroics can be model systems for studying fundamental laws of nature that have analogous mathematical descriptions.

I close, by way of an outlook, by mentioning two examples of where the concepts and formalisms that have been developed or refined through the study of multiferroics and magnetoelectrics in the last decades have begun to permeate seemingly unrelated areas of physics \cite{Fiebig_et_al:2016}, and which capture my ongoing interest in multiferroics beyond the electric-field control of magnetism.

Ferroic ordering in ferromagnets and ferroelectrics is a well-established concept to describe the associated macroscopic thermodynamic phenomena of magnetization and polarization and their associated time-reversal and space-inversion symmetry breakings respectively. In some magnetoelectric materials, however, both symmetries are broken without the onset of magnetization or polarization suggesting a new kind of {\it hidden order}. Here, the magnetoelectric monopole, quadrupole and toroidal moment described above are candidates for the ordered quantity, and there is considerable activity in trying to directly measure these entities \cite{Staub_et_al:2009,Lovesey/Khalyavin/Staub:2015}. The latter has even been proposed as a possible candidate for the hidden order parameter in the pseudo-gap phase of the high-$T_c$ cuprate superconductors \cite{Fechner_et_al:2016}. These magnetoelectric multipoles provide a link to the rapidly growing field of topological insulators, which have traditionally been characterized in terms of their surface conductance, but which can be recast in the formalism of bulk magnetoelectric behavior with a quantized diagonal and isotroptic magnetoelectric response \cite{Armitage/Wu:2019}. Their occurrence in metals, while prohibiting a conventional magnetoelectric response, provides a link to the rapidly growing field of antiferromagnetic spintronics \cite{Thole/Spaldin:2018}.

If a ferroic phase transition can be suppressed to zero-temperature using a variable such as pressure or applied field, then the ordering is determined by quantum rather than thermal effects, resulting in exotic behavior even at finite temperatures. Such quantum criticality is well established for magnetic ordering, where it is believed to be responsible for example for heavy-fermion superconductivity, and has also recently been identified in ferroelectrics \cite{Rowley_et_al:2014}, where it is believed to be associated with the unconventional superconductivity in SrTiO$_3$ \cite{Edge_et_al:2015,Stucky_et_al:2016}. The concept of multiferroic quantum criticality (Fig.~\ref{fig:MFQC}), in which both magnetic and ferroelectric quantum criticality occur in the same system, was recently introduced \cite{Narayan_et_al:2019} and candidate materials that could host the phenomenon were suggested. Experimental signatures should include modified and coupled scaling relations of magnetic and dielectric susceptibilities  \cite{Narayan_et_al:2019}; what new physics will arise remains an intriguing open question.

In a report on the development of the designer multiferroic to search for the electric dipole moment of the electron, National Geographic News headlined with {\it Universe's Existence May Be Explained by New Material} \cite{Roach:2010}. While this might be overstating the case somewhat, it is clear that the reach of multiferroics has far exceeded early expectations, and that multiferroic concepts will continue to influence the thinking of physicists for many years to come.

\newpage

\end{document}